\documentclass[journal]{IEEEtran}

\usepackage[dvips]{color}
\usepackage{graphicx}
\usepackage{amsmath}

\begin{document}

\title{Security-Reliability Tradeoff for Distributed Antenna Systems in Heterogeneous Cellular Networks}

\markboth{IEEE Transactions on Wireless Communications (ACCEPTED TO APPEAR)}%
{Yulong~Zou \emph{et al.}: Security-Reliability Tradeoff for Distributed Antenna Systems in Heterogeneous Cellular Networks}

\author{Yulong~Zou,~\IEEEmembership{Senior Member,~IEEE}, Ming~Sun, Jia~Zhu, and Haiyan~Guo

\thanks{Manuscript received November 11, 2017; revised February 11, 2018, June 11, 2018, and October 2, 2018; accepted October 17, 2018. The editor coordinating the review of this paper and approving it for publication was Prof. S. Dey.}

\thanks{This work was partially supported by the National Natural Science Foundation of China (Grant Nos. 61522109 and 91738201), the Natural Science Foundation of Jiangsu Province (Grant Nos. BK20150040 and BK20171446), Open Research Fund of the Key Laboratory of Ministry of Education in Broadband Wireless Communication and Sensor Network Technology (Grant No. JZNY201707), and the 1311 Talent Program of Nanjing University of Posts and Telecommunications.}

\thanks{The authors with the School of Telecommunications and Information Engineering, Nanjing University of Posts and Telecommunications, Nanjing 210003, Jiangsu, P. R. China. (Email: \{yulong.zou\}@njupt.edu.cn)}

\thanks{Corresponding author is Yulong Zou.}

}

\maketitle

\begin{abstract}
In this paper, we investigate physical-layer security for a spectrum-sharing heterogeneous cellular network comprised of a macro cell and a small cell, where a passive eavesdropper is assumed to tap the transmissions of both the macro cell and small cell. In the macro cell, a macro base station (MBS) equipped with multiple distributed antennas sends its confidential information to a macro user (MU) through an opportunistic transmit antenna. Meanwhile, in the small cell, a small base station (SBS) transmits to a small user (SU) over the same spectrum used by MBS. We propose an interference-canceled opportunistic antenna selection (IC-OAS) scheme to enhance physical-layer security for the heterogeneous network. To be specific, when MBS sends its confidential message to MU through an opportunistic distributed antenna, a special signal is artificially designed and emitted at MBS to ensure that the received interference at MU from SBS is canceled out. For comparison, the conventional interference-limited opportunistic antenna selection (IL-OAS) is considered as a benchmark. We characterize the security-reliability tradeoff (SRT) for the proposed IC-OAS and conventional IL-OAS schemes in terms of deriving their closed-form expressions of intercept probability and outage probability. Numerical results show that compared with the conventional IL-OAS, the proposed IC-OAS scheme not only brings SRT benefits to the macro cell, but also has the potential of improving the SRT of small cell by increasing the number of distributed antennas. Additionally, by jointly taking into account the macro cell and small cell, an overall SRT of the proposed IC-OAS scheme is shown to be significantly better than that of the conventional IL-OAS approach in terms of a sum intercept probability versus sum outage probability.

\end{abstract}


\section{Introduction}
\IEEEPARstart{I}{n} order to address an explosive increase in data traffic generated by various wireless devices (e.g., smart phones, tablets and laptops) [1], [2], heterogeneous networks (HetNets) are emerging as an effective paradigm to enhance the system capacity and coverage for guaranteeing the quality-of-service (QoS) of subscribers [3]-[6]. HetNets are usually composed of various macro cells, small cells (e.g., pico cells and femto cells), and relay stations, where low-power small cells (ranging from 250mW to 2W) are underlaid in higher-power macro cells (5W-40W) [4]. Typically, macro base stations (MBSs) and small base stations (SBSs) are permitted to simultaneously transmit their respective confidential messages over the same spectrum band. As a result, the spectral efficiency can be significantly improved along with an increased network capacity [7], [8]. However, mutual interference may exist among the macro cells and small cells, as the same spectrum band is simultaneously accessed in an underlay manner. In order to alleviate the mutual interference problem, an interference-aware muting scheme was proposed in [9] to reduce the interference level below a tolerable threshold. In [10], the authors proposed an interference cancelation scheme at MBS to cancel out the cross-tier interference received at a small-cell subscriber and derived a closed-form outage probability expression of HetNets. In [11]-[13], the authors explored interference management for the sake of improving the network coverage of HetNets.

However, due to the broadcast nature of wireless communications [14] and the open system architecture of HetNets [15], confidential messages transmitted to legitimate users are extremely vulnerable to eavesdropping attacks. Thus, it is of importance to investigate the transmission confidentiality of HetNets against eavesdropping. Traditionally, key-based cryptographic methods were employed to guarantee the confidentiality of wireless transmissions. However, with the fast development of computing technology, the eavesdropper may have a sufficiently high computing power to crack the secret key. Since the first physical-layer security work carried out by Wyner in [16], where the secrecy capacity was given as the difference between the capacity of main channel and that of wiretap channel, an increasing research attention has been paid to this research field, which is considered as a promising means of achieving a perfect secrecy against eavesdropping. During the past decades, cooperative relay [17]-[19], beamforming [20]-[22], and multiuser scheduling [23]-[25] were proposed to strengthen the physical-layer security for different wireless network scenarios. Moreover, distributed multiple-input multiple-output (MIMO) systems were also investigated from the physical-layer security perspective in [26] and [27].

To the best of our knowledge, most of existing research efforts have been focused on the network coverage [28], [29], energy efficiency [30], [31], and spectral efficiency [32], [33] of HetNets. Besides, there also exits some research work on physical-layer security for spectrum-sharing HetNets [34]-[36]. Typically, cognitive radio (CR) networks can be envisioned as one type of spectrum-sharing HetNets. In CR systems, an unlicensed secondary user is allowed to access the licensed spectrum that is not used by a primary user, where the primary user has a higher priority than the secondary user in accessing the spectrum. Moreover, the primary user and secondary user belong to two different networks, which are typically separated and independent from each other. In [18] and [37], the authors investigated physical-layer security of secondary transmissions without affecting the QoS of primary transmissions for CR networks. In [38], the authors studied the secrecy-optimized resource allocation for device-to-device communication systems. In [39], a secrecy coverage probability was derived in downlink MIMO multi-hop HetNets. It is noted that mutual interference between the macro cells and small cells is critical in underlay HetNets, which was intelligently exploited in [1] to defend against eavesdropping for spectrum-sharing HetNets. In [1], an interference-canceled underlay spectrum sharing (IC-USS) scheme was proposed for canceling out the interference received at a macro user (MU) while interfering with an unintended eavesdropper.

Differing from the system model with a single antenna as studied in [1], we consider multiple distributed antennas available in the macro cell of heterogeneous cellular networks to guarantee the QoS of far-off subscribers [40], [41]. Both MBS and SBS are connected to a core network via fiber cables, e.g., a mobile switch center (MSC) in the global system for mobile communication (GSM) and a mobility management entity (MME) in the long term evolution (LTE) [1], which ensures the real-time interaction between MBS and SBS. The main contributions of this paper are summarized as follows. First, combining the interference cancelation of [1] and opportunistic antenna selection (OAS) techniques, we propose an interference-canceled OAS (IC-OAS) scheme for the sake of improving the security-reliability tradeoff (SRT) performance of heterogeneous cellular networks. The proposed IC-OAS is different from the zero-forcing beamforming of [42], where multiple transmit antennas are employed to emit a source signal simultaneously with a beamforming vector, which requires complex symbol-level synchronization between the multiple antennas for avoiding severe inter-symbol interference. By contrast, in our IC-OAS scheme, only a single distributed antenna is chosen to transmit the source signal, which reduces the complexity of distributed antenna synchronization. Second, we derive closed-form expressions of intercept probability and outage probability for the proposed IC-OAS as well as conventional interference-limited OAS (IL-OAS) schemes. {{Numerical results show that the proposed IC-OAS scheme is capable of improving the SRTs of both the macro cell and small cell, as compared to the conventional IL-OAS approach. Additionally, a normalized sum of intercept probability and outage probability (denoted IOP for short) of both the macro cell and small cell versus a ratio of the transmit power of SBS to that of MBS, referred to as the small-to-macro ratio (SMR), is evaluated for the IL-OAS and IC-OAS schemes. It is demonstrated that the normalized sum IOP of our IC-OAS scheme can be further optimized with regard to the SMR and the optimized sum IOP of proposed IC-OAS is much better than that of conventional IL-OAS.}}

The reminder of this paper is organized as follows. In Section II, we present the system model of a spectrum-sharing heterogeneous cellular network and propose the IC-OAS scheme. For comparison purposes, the conventional IL-OAS scheme is also presented. In Section III, we characterize the SRT for both IC-OAS and IL-OAS in terms of deriving their closed-form expressions of intercept probability and outage probability. Next, numerical SRT results and discussions are provided in Section IV. Finally, some concluding remarks are given in Section V.

\section{Spectrum-sharing Heterogeneous Cellular Networks}
In this section, we first present the system model of a heterogeneous cellular network, where a macro cell coexists with a small cell and an eavesdropper is assumed to tap legitimate transmissions of both the macro cell and small cell. Next, an underlay spectrum sharing (USS) mechanism [1] is considered for the heterogeneous cellular network.

\subsection{System Model}

\begin{figure}
  \centering
  {\includegraphics[scale=0.5]{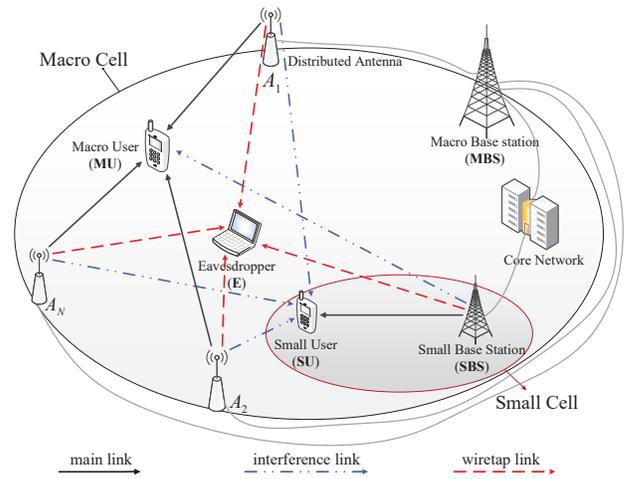}\\
  \caption{A heterogeneous cellular network comprised of a macro cell and a small cell in the presence of an eavesdropper.}\label{Fig1}}
\end{figure}
Fig. 1 shows a heterogeneous cellular network composed of a macro cell and a small cell. Differing from the separated independent primary and secondary networks in CR systems, the macro cell and small cell are coordinated via the core network in heterogeneous cellular networks, through which the reliable information exchange can be achieved between MBS and SBS. This guarantees that a specially-designed signal becomes possible at MBS, since the design of such a special signal requires the reliable exchange of some system information between MBS and SBS [1], e.g., the channel state information (CSI), transmit power, and so on. Moreover, although only a single small cell is taken into account in this paper, a possible extension can be considered for a large-scale heterogeneous network consisting of massive small cells with the help of stochastic geometry [42] and user scheduling [45]. Additionally, if more than one SBS is available, the given spectrum may be divided into multiple orthogonal sub-bands which are then allocated to different SBSs. In this way, only one SBS is assigned to simultaneously access an orthogonal sub-band with MBS for the sake of alleviating the complex synchronization among spatially-distributed SBSs.

In the macro cell, MBS first sends its confidential message to distributed antennas ${A_i}$ ($i = 1,2, \cdots ,N$), where $N$ is the number of distributed antennas. Then, a single antenna is opportunistically selected to transmit the confidential message of MBS to MU. Meanwhile, in the small cell, SBS transmits its signal to a small user (SU) over the same spectrum used by MBS. Moreover, a passive eavesdropper is assumed to tap ${A_i}$-MU and SBS-SU transmissions. To improve the spectrum utilization, we consider an USS mechanism for the heterogeneous cellular network throughout this paper. Specifically, in the USS mechanism, MBS and SBS are permitted to simultaneously transmit their respective confidential messages over the same spectrum band. However, in order to guarantee the QoS of heterogeneous cellular networks, the transmit powers of MBS and SBS should be controlled to limit mutual interference. For notational convenience, let $P_M$ and $P_S$ denote the transmit powers of MBS and SBS, respectively. Moreover, an additive white Gaussian noise (AWGN) is encountered at any receiver of Fig. 1 with a zero mean and a variance of ${N_0}$.

\subsection{Conventional IL-OAS}
In this section, we present the conventional IL-OAS scheme as a baseline, where MBS and SBS are allowed to simultaneously access the same spectrum band. In order to guarantee the QoS of macro cell, the transmit power of SBS is controlled for limiting the interference to macro cell [1]. For the macro cell, MBS first transmits its confidential message ${x_M}$ ($E( {{{| {{x_M}} |}^2}} = 1$) to ${A_i}$ through a fiber-optic cable. Then, a single antenna is opportunistically selected to forward its received messages to MU at a power of ${P_M}$. By contrast, in the small cell, SBS directly transmits its message $x_S$ ($E( {{{| {{x_S}} |}^2}} ) = 1$) to SU over the same spectrum used by MBS at a power of $P_S$. The aforementioned transmission process leads to the fact that a mixed signal of ${x_M}$ and ${x_S}$ is received at MU and SU. For notational convenience, let ${\cal{A}}  = \left\{ {{A_i}| {i = 1,2, \cdots ,N} } \right\}$ represent the set of $N$ distributed antennas.

For the macro cell, if a distributed antenna $A_i$ is selected to transmit the signal $x_M$, the received signal at MU can be expressed as
\begin{equation}\label{equa1}
y_{{m}}^{{\textrm{IL}}} = {h_{A_im}}\sqrt {{P_M}} {x_M} + {h_{Sm}}\sqrt {{P_S}} {x_{S}} + {n_{m}},
\end{equation}
where ${h_{{A_i}m}} = \sqrt {d_{{A_i}m}^{ - \alpha_{A_im} }} {g_{{A_i}m}}$, ${h_{Sm}} = \sqrt {d_{Sm}^{ - \alpha_{Sm} }}{g_{Sm}}$, $g_{{A_i}m}$ and $g_{Sm}$ represent the small-scale fading gains of $A_i$-MU and SBS-MU channels, respectively, $d_{{A_i}m}$ and $d_{Sm}$ are the distances of $A_i$-MU and SBS-MU transmissions, $\alpha_{A_im}$ and $\alpha_{Sm}$ are path loss factors of the $A_i$-MU and SBS-MU channels, respectively, and ${n_{m}}$ is the AWGN encountered at MU. According to Shannon's capacity formula, we can obtain the channel capacity of $A_i$-MU from (1) as
\begin{equation}\label{equa2}
{C_{{A_i}m}^{{\textrm{IL}}} = {\log _2}(1 + \frac{{{\gamma _M}{{| {{h_{{A_i}m}}} |}^2}}}{{{\gamma _S}{{| {{h_{Sm}}} |}^2} + 1}})},
\end{equation}
where ${\gamma _M} = {P_M}/{N_0}$ and ${\gamma _S} = {P_S}/{N_0}$ are the signal-to-noise ratios (SNRs) of MBS and SBS, respectively. Typically, the antenna $A_i$ with the highest instantaneous channel capacity of $C_{{A_i}m}^{{\textrm{IL}}}$ is selected to assist the MBS-MU transmission. Thus, from (2), an opportunistic antenna selection criterion is given by
\begin{equation}\label{equa3}
{A = {\kern 1pt} {\kern 1pt} {\rm{arg}}\mathop {{\rm{max}}}\limits_{{A_i} \in {\cal{A}} } C_{{A_i}m}^{{\textrm{IL}}} = {\kern 1pt} {\kern 1pt} {\rm{arg}} \mathop {\max }\limits_{{A_i} \in {\cal{A}} } {{{| {{h_{{A_i}m}}} |}^2}}},
\end{equation}
which shows that the CSI ${ {{h_{A_im}}} }$ is used to perform the opportunistic antenna selection. According to (3), the channel capacity of MBS-MU is obtained as
\begin{equation}\label{equa4}
\begin{split}
C_{Mm}^{{\textrm{IL}}}& = {\log _2}(1 + \mathop {\max }\limits_{{A_i} \in {\cal{A}} } \frac{{{\gamma _M}{{| {{h_{{A_i}m}}} |}^2}}}{{{\gamma _S}{{| {{h_{Sm}}} |}^2} + 1}})\\
& = {\log _2}(1 + \frac{{{\gamma _M}{{| {{h_{Am}}} |}^2}}}{{{\gamma _S}{{| {{h_{Sm}}} |}^2} + 1}}) ,
\end{split}
\end{equation}
where subscript $A$ denotes the distributed antenna selected. Also, for the small cell, the received signal at SU can be similarly expressed as
\begin{equation}\label{equa5}
{y_{s}^{{\textrm{IL}}} = {h_{Ss}}\sqrt {{P_S}} {x_S} + {h_{As}}\sqrt {{P_M}} {x_{M}} + {n_s}},
\end{equation}
where ${h_{Ss}} = \sqrt {d_{Ss}^{ - \alpha_{Ss} }} {g_{Ss}}$, ${h_{As}} = \sqrt {d_{As}^{ - \alpha_{As} }}{g_{As}}$, $g_{Ss}$ and $g_{As}$ denote the small-scale fading gains of SBS-SU and $A$-SU channels, respectively, $d_{Ss}$ and $d_{As}$ are the distances of SBS-SU and $A$-SU transmissions, $\alpha_{Ss}$ and $\alpha_{As}$ are path loss factors of the SBS-SU and $A$-SU channels, respectively, and ${n_s}$ is the AWGN encountered at SU. Similarly, the channel capacity of SBS-SU is obtained from (5) as
\begin{equation}\label{equa6}
{C_{Ss}^{{\textrm{IL}}} = {\log _2}(1 + \frac{{{\gamma _S}{{| {{h_{Ss}}} |}^2}}}{{{\gamma _M}{{| {{h_{As}}} |}^2}}+1})}.
\end{equation}

Meanwhile, the eavesdropper may overhear both the MBS-MU and SBS-SU transmissions. As a result, the corresponding received signal at the eavesdropper can be written as
\begin{equation}\label{equa7}
{y_e^{{\textrm{IL}}} = {h_{Ae}}\sqrt {{P_M}} {x_M} + {h_{Se}}\sqrt {{P_S}} {x_S} + {n_e}},
\end{equation}
where ${h_{Ae}} = \sqrt {d_{Ae}^{ - \alpha_{Ae} }} {g_{Ae}}$, ${h_{Se}} = \sqrt {d_{Se}^{ - \alpha_{Se} }}{g_{Se}}$, $g_{Ae}$ and $g_{Se}$ represent the small-scale fading gains of $A$-E and SBS-E channels, respectively, $d_{Ae}$ and $d_{Se}$ are the distances of $A$-E and SBS-E transmissions, $\alpha_{Ae}$ and $\alpha_{Se}$ are path loss factors of the $A$-E and SBS-E channels, respectively, and ${n_e}$ is the AWGN encountered at the eavesdropper. For simplicity, we here assume that the eavesdropper decodes $x_M$ and $x_S$ separately without the help of successive interference cancelation. Based on the Shannon's capacity formula, the channel capacity of MBS-E and that of SBS-E are given by
\begin{equation}\label{equa8}
{C_{Me}^{{\textrm{IL}}} = {\log _2}( {1 + \frac{{{\gamma _M}{{| {{h_{Ae}}} |}^2}}}{{{\gamma _S}{{| {{h_{Se}}} |}^2} + 1}}} )},
\end{equation}
and
\begin{equation}\label{equa9}
{C_{Se}^{{\textrm{IL}}} = {\log _2}(1 + \frac{{{\gamma _S}{{| {{h_{Se}}} |}^2}}}{{{\gamma _M}{{| {{h_{Ae}}} |}^2} + 1}})}.
\end{equation}

\subsection{Proposed IC-OAS}
In this section, we propose an IC-OAS scheme, where MBS and SBS are also permitted to access the same spectrum simultaneously, leading to an existence of mutual interference between the macro cell and small cell, as aforementioned. {{For the sake of canceling out the interference received at MU from SBS, a special signal denoted by ${x_{A_i}}$ is designed and emitted through a selected antenna $A_i$ at MBS. When a mixed signal of $x_M$ and $x_{A_i}$ is transmitted at MBS, a weight coefficient $w_S$ is utilized at SBS for transmitting its signal $x_S$ at a power of $P_S$. The instantaneous and average transmit powers of $x_{A_i}$ are represented by $P_{A_i}$ and $\bar P_{A_i}$, respectively. For a fair comparison with the IL-OAS scheme, the total average transmit power of ${x_M}$ and ${x_{A_i}}$ is constrained to ${P_M}$ at MBS. In this sense, the transmit power of ${x_M}$ is given by ${P_M}-{\bar P_{A_i}}$. Obviously, the average transmit power of $x_{A_i}$ should satisfy the following inequality}}
\begin{equation}\label{equa10}
{0 \le {\bar P_{A_i}} \le {P_M}.}
\end{equation}
Considering that a distributed antenna $A_i$ is selected to transmit the mixed signal of ${x_M}$ and ${x_{A_i}}$, we can express the received signal at MU as
\begin{equation}\label{equa11}
{\begin{split}
y_m^{{\textrm{IC}}} = & {h_{{A_i}m}}(\sqrt {{P_M} - {\bar P_{A_i}}} {x_M} + {x_{A_i}}) \\
&+ {h_{Sm}}\sqrt {{P_S}} {w_S x_S} + {n_m},\\
\end{split}}
\end{equation}
where $h_{A_im}$ represents a fading coefficient of the channel from the distributed antenna $A_i$ to MU. For the sake of neutralizing the interference term of (11), the following equality should be satisfied
\begin{equation}\nonumber
{h_{{A_i}m}}{x_{A_i}} + {h_{Sm}}\sqrt {{P_S}} {w_S x_S}= 0,
\end{equation}
from which various solutions of $[x_{A_i}, w_S]$ can be found for the interference neutralization. Throughout this paper, a solution of $[x_{A_i}, w_S]$ to the preceding equation is given by
\begin{equation}\label{equa12}
[x_{A_i}, w_S] = \frac{1}{{{\sigma _{A_im}}}}[- |{h_{Sm}}|{e^{ - j{\theta _{A_im}}}}{\sqrt {{P_s}} }{x_S},|{h_{A_im}}|{e^{ - j{\theta _{Sm}}}}],
\end{equation}
where $\sigma_{A_im}^2 = E(|{h_{A_im}}{|^2})$ represents the variance of the channel from the distributed antenna $A_i$ to MU, ${{\theta _{A_im}}}$ and ${{\theta _{Sm}}}$ denote the phase of the channel from the distributed antenna $A_i$ to MU and that from SBS to MU, respectively. It can be observed from (12) that the design of $[{x_{A_i}}, w_S]$ requires the knowledge of ${h_{A_im}}$, ${h_{Sm}}$, $\sigma_{A_im}^2$, $P_S$ and ${x_S}$ at MBS and SBS. Typically, the CSIs of $h_{A_im}$ and $h_{Sm}$ are usually estimated at MU and then fed back to MBS and SBS [43]. The statistical CSI of $\sigma_{A_im}^2$ can be readily obtained by exploiting the accumulated knowledge of instantaneous CSIs of $h_{A_im}$. Moreover, the information of $x_S$ and $P_S$ may be acquired at MBS through the core network. {{It is worth mentioning that the message $x_S$ is not generated at SBS, which is typically initiated by another user terminal of cellular networks and sent via the core network first to SBS that then forwards to SU through its air interface in the subsequent stage. Thus, when the core network sends the message $x_S$ to SBS in the first stage, the same copy of $x_S$ can be received and stored at MBS simultaneously. This guarantees that no significant amount of extra time delay is incurred at MBS in obtaining $x_S$ as compared to SBS, regardless of the latency of the core network. Additionally, a small cell is generally deployed for various indoor scenarios with narrow coverage, where user terminals often stay stationary or move at a very low speed ($0$-$3$km/h) [46]. In this case, the transmission distance of SBS-SU is normally stationary along with a quasi-static path loss and thus the transmit power of SBS $P_S$ is stable, which can be pre-determined before the information transmission and sent to MBS in advance. Therefore, the information of both $x_S$ and $P_S$ can be pre-acquired at MBS before starting the transmission of $x_M$ and $x_S$, implying that our interference cancelation mechanism is nonsensitive to the time delay of the core network. It is of particular interest to examine the impact of channel estimation errors and feedback delay on the SRT performance of our IC-OAS scheme, which is considered for further work. From (12), the instantaneous and average transmit powers of $x_{A_i}$ are given by
\begin{equation}\label{equa13}
[{P_{A_i}}, \bar P_{A_i}] = [ \frac{{|{h_{Sm}}{|^2}}}{{\sigma _{A_im}^2}}{P_S} ,\frac{{\sigma _{Sm}^2}}{{\sigma _{A_im}^2}}{P_S}],
\end{equation}
where $\sigma _{Sm}^2 $ and $\sigma _{A_im}^2$ are the means of ${| {{h_{Sm}}} |^2}$ and ${| {{h_{A_im}}} |^2}$, respectively. Combining (10) and (13), we obtain
\begin{equation}\label{equa14}
{\frac{{{P_M}}}{{{P_S}}} > \frac{{\sigma _{Sm}^2}}{{\sigma _{A_im}^2}},}
\end{equation}
which indicates that the interference received at MU from SBS can be perfectly canceled out when the average received signal strength from MBS is stronger than the one from SBS. It needs to be pointed that there may exist an optimal solution of $[{x_{A_i}}, w_S] $ in terms of maximizing the secrecy performance of MBS-MU transmissions, which is out of the scope and may be considered for future work. Substituting (12) into (11) yields
\begin{equation}\label{equa15}
{y_{m}^{\textrm{IC}} = {h_{A_im}} {\sqrt {{P_M} - {\bar P_{A_i}}} {x_M}}  + {n_m},}
\end{equation}
from which the capacity of the channel from a distributed antenna $A_i$ to MU is given by
\begin{equation}\label{equa16}
C_{{A_i}m}^{{\textrm{IC}}} = {\log _2}[1 + ({\gamma _M} - \frac{{\sigma _{Sm}^2}}{{\sigma _{{A_i}m}^2}}{\gamma _S}){| {{h_{{A_i}m}}} |^2}].
\end{equation}
Typically, the distributed antenna $A_i$ with the highest instantaneous channel capacity of $C_{{A_i}m}^{{\textrm{IC}}}$ is selected to transmit the MBS' signal. Thus, from (16), an opportunistic antenna selection criterion is expressed as
\begin{equation}\label{equa17}
A  =  {\rm{arg}} \mathop {\max }\limits_{{A_i} \in {\cal{A}} } ({\gamma_M} - \frac{{\sigma _{Sm}^2}}{{\sigma _{{A_i}m}^2}}{\gamma_S}){| {{h_{{A_i}m}}} |^2},
\end{equation}
where the subscript `$A$' denotes the distributed antenna selected. Moreover, when the channel fading coefficients $|h_{A_im}|^2$ for different distributed antennas are considered to be independent identically distributed (i.i.d.), the aforementioned antenna selection criterion of (17) becomes the same as the conventional one of (3). Hence, the channel capacity of MBS-MU relying on the opportunistic antenna selection of (17) is obtained as
\begin{equation}\label{equa18}
{\begin{split}
C_{Mm}^{{\textrm{IC}}} & = {\log _2}[1 + \mathop {{\rm{max}}}\limits_{{A_i} \in {\cal{A}} } ({\gamma _M} - \frac{{\sigma _{Sm}^2}}{{\sigma _{{A_i}m}^2}}{\gamma _S}){| {{h_{{A_i}m}}} |^2}].\\
\end{split}}
\end{equation}
Also, for the small cell, the received signal at SU can be similarly written as
\begin{equation}\label{equa19}
y_s^{{\textrm{IC}}} = {h_{Ss}}\sqrt {{P_S}} {w_Sx_S} + {h_{As}}(\sqrt {{P_M} - {\bar P_{A}}} {x_M} + {x_{A}}) + {n_s},
\end{equation}
where ${h_{As}}$ represents a fading coefficient of the channel from the selected antenna $A$ to SU, $x_A$ denotes the specially-designed signal emitted at the selected antenna $A$ and $\bar P_A$ is the average transmit power of $x_A$. It can be observed from (19) that although the term ${h_{As}}x_{A}$ contains the SBS' signal $x_S$ as implied from (12), it is not aligned and thus interfered with ${h_{Ss}}\sqrt {{P_S}} w_S {x_S}$, since the signal $x_{A}$ is designed to be neutralized with the interference received at MU. Moreover, an advanced signal processing technique e.g. selection diversity combining (SDC) may be employed at the SU receiver by jointly exploiting the terms ${h_{As}}x_{A}$ and ${h_{Ss}}\sqrt {{P_S}} w_S {x_S}$ for decoding $x_S$, which can be also adopted by the eavesdropper, thus no improvement is expected for the small cell from an SRT perspective. For simplicity, the signal $x_{A}$ is treated as an interference at both the SU and eavesdropper in decoding $x_S$. Hence, the capacity of SBS-SU channel can be obtained from (12) and (19) as
\begin{equation}\label{equa20}
C_{Ss}^{{\textrm{IC}}} = {\log _2}[1 + \frac{{{\gamma _S}{{| {{h_{Ss}}} |}^2{| {{h_{Am}}} |}^2/{{{\sigma_{Am}^2}} }}}}{{(\gamma _M+\gamma_S g_{Sm} ){{| {{h_{As}}} |}^2}}+1}],
\end{equation}
where $g_{Sm}=(|h_{Sm}|^2-\sigma^2_{Sm})/{\sigma_{Am}^2}$. Meanwhile, the eavesdropper is considered to tap both the MBS-MU and SBS-SU transmissions. As a result, the corresponding received signal at the eavesdropper can be written as
\begin{equation}\label{equa21}
{y_e^{{\textrm{IC}}} = {h_{Ae}}(\sqrt {{P_M} - {\bar P_{A}}} {x_M} + {x_{A}}) + {h_{Se}}\sqrt {{P_S}} {w_S x_S} + {n_e}},
\end{equation}
where ${h_{Ae}}$ represents a fading coefficient of the channel from the selected antenna $A$ to the eavesdropper. Again, considering that the eavesdropper decodes $x_M$ and $x_S$ separately without successive interference cancelation as well as using (12) and (13), we can obtain the channel capacity of MBS-E and that of SBS-SU as
\begin{equation}\label{equa22}
C_{Me}^{{\textrm{IC}}} = {\log _2}[1 + \frac{{({\gamma _M}{{\sigma _{Am}^2}} - {\gamma _S}{{\sigma _{Sm}^2}}){{| {{h_{Ae}}} |}^2}}}{{{\gamma _S}({{| {{h_{Ae}}} |}^2}{|{h_{Sm}|^2}} + {{| {{h_{Se}}} |}^2}{|{h_{Am}|^2}}) + {{\sigma _{Am}^2}}}}],
\end{equation}
and
\begin{equation}\label{equa23}
C_{Se}^{{\textrm{IC}}} = {\log _2}[1 + \frac{{{\gamma _S}{{| {{h_{Se}}} |}^2}}{|{h_{Am}|^2}}/{{\sigma_{Am}^2}}}{{({\gamma _M} + \gamma_S g_{Sm}){{| {{h_{Ae}}} |}^2} + 1}}],
\end{equation}
where $g_{Sm}=(|h_{Sm}|^2-\sigma^2_{Sm})/{\sigma_{Am}^2}$.

\section{Security and Reliability Performance Analysis}
In this section, we characterize the SRT of proposed IC-OAS and conventional IL-OAS schemes in terms of deriving their closed-form expressions of intercept probability and outage probability over Rayleigh fading channels. Following [44] and [47], an outage probability of legitimate transmissions is given by
\begin{equation}\label{equa24}
{{P_{\text {out}}} = \Pr ( {{C_m} < {R_o}} ),}
\end{equation}
where $C_m$ denotes the channel capacity of legitimate transmissions and ${R_o}$ is an overall transmission rate. Moreover, an intercept probability can be written as
\begin{equation}\label{equa25}
{{P_{\text {int}}} = \Pr ( {{C_e} > {R_o} - {R_s}} ),}
\end{equation}
where $C_e$ represents the wiretap channel capacity and $R_s$ is a secrecy rate. It can be observed from (25) that when the wiretap channel capacity ${C_e}$ becomes higher than the rate difference of ${R_o - R_s}$, a prefect secrecy is impossible and an intercept event happens in this case.

\subsection{Conventional IL-OAS}
In this subsection, we analyze the outage probability and intercept probability of the macro-cell and small-cell transmissions relying on the conventional IL-OAS scheme. From (24), an outage probability of the MBS-MU transmission is written as
\begin{equation}\label{equa26}
{P_{M\textrm{out}}^{{\textrm{IL}}} = \Pr ( {C_{{Mm}}^{{\textrm{IL}}} < R_M^o} ),}
\end{equation}
where $R_M^o$ is an overall data rate of MBS-MU transmission. Substituting $C_{{Mm}}^{{\textrm{IL}}}$ from (4) into (26) yields
\begin{equation}\label{equa27}
{P_{{M\textrm{out}}}^{{\textrm{IL}}} = \Pr (\mathop {\max }\limits_{{A_i} \in {\cal{A}} } \frac{{{{| {{h_{{A_i}m}}} |}^2}}}{{{\gamma _S}{{| {{h_{Sm}}} |}^2} + 1}} < {\Delta _M})},
\end{equation}
where $\Delta_M  = ( {{2^{R_M^o}} - 1})/{\gamma _M}$. Proceeding as in Appendix A, we can obtain $P_{{M\textrm{out}}}^{{\textrm{IL}}}$ as
\begin{equation}\label{equa28}
\begin{split}
P_{M{\textrm{out}}}^{{\textrm{IL}}} = &1 + \sum\limits_{j = 1}^{{2^{|{\cal{A}} |}} - 1} {{{( - 1)}^{|A(j)|}}} \exp ( - \sum\limits_{{A_i} \in {\cal{A}}(j)} {\frac{{{\Delta _M}}}{{\sigma _{{A_i}m}^2}}} )\\
&\quad\quad\quad\quad\quad\times{(1 + \sum\limits_{{A_i} \in {\cal{A}}(j)} {\frac{{{\Delta _M}{\gamma _S}\sigma _{Sm}^2}}{{\sigma _{{A_i}m}^2}}} )^{ - 1}},
\end{split}
\end{equation}
where ${A(j)}$ represents the $j$-th non-empty subset of the antenna set ${\cal{A}} $. Similarly, by using (6) and (24), the outage probability of SBS-SU transmission is expressed as
\begin{equation}\label{equa29}
{P_{S\textrm{out}}^{{\textrm{IL}}} = \Pr ( {C_{{Ss}}^{{\textrm{IL}}} < R_S^o} ),}
\end{equation}
where $R_S^o$ is the overall data rate of SBS-SU transmission. Substituting $C_{{Ss}}^{{\textrm{IL}}}$ from (6) into (29) yields
\begin{equation}\label{equa30}
{\begin{split}
P_{S{\textrm{out}}}^{{\textrm{IL}}}
&= \Pr (\frac{{{{| {{h_{Ss}}} |}^2}}}{{{\gamma _M}{{| {{h_{As}}} |}^2} + 1}} < {\Delta _S})\\
& = \sum\limits_{{A_i} \in {\cal{A}} } {\Pr \left( \begin{split}
&\frac{{{{| {{h_{Ss}}} |}^2}}}{{{\gamma _M}{{| {{h_{{A_i}s}}} |}^2} + 1}} < {\Delta _S},\\
&\mathop {\max}\limits_{{A_k} \in {\cal{A}}  - \left\{ {{A_i}} \right\}} {{{{| {{h_{{A_k}m}}} |}^2}}} < {{{{| {{h_{{A_i}m}}} |}^2}}}
\end{split} \right)},
\end{split}}
\end{equation}
where $\Delta_S  = ( {{2^{R_S^o}} - 1})/{\gamma _S}$. Since ${{{| {{h_{Ss}}} |}^2}}$, ${{{| {{h_{A_is}}} |}^2}}$ and ${{{| {{h_{A_im}}} |}^2}}$ are independent exponentially distributed random variables with respective means of $\sigma _{Ss}^2$, $\sigma _{A_is}^2$ and $\sigma _{A_im}^2$, we can further obtain $P_{{Ss\textrm{-out}}}^{{\textrm{IL}}}$ as
\begin{equation}\label{equa31}
P_{S{\textrm{out}}}^{{\textrm{IL}}} = \sum\limits_{{A_i} \in {\cal{A}} } {{P_{S{\textrm{out},A_i}}^{{\textrm{IL}}}}{P(A_i)}},
\end{equation}
where the terms ${P_{S{\textrm{out},A_i}}^{{\textrm{IL}}}} $ and ${P(A_i)}$ are given by
\begin{equation}\label{equa32}
\begin{split}
{P_{S{\textrm{out},A_i}}^{{\textrm{IL}}}} &= {\Pr (\frac{{{{| {{h_{Ss}}} |}^2}}}{{{\gamma _M}{{| {{h_{{A_i}s}}} |}^2} + 1}} < {\Delta _S})}\\
&= 1 - \frac{{\sigma _{Ss}^2}}{{\sigma _{Ss}^2 + {\gamma _M}\sigma _{{A_i}s}^2{\Delta _S}}}\exp ( - \frac{{{\Delta _S}}}{{\sigma _{Ss}^2}}),
\end{split}
\end{equation}
and
\begin{equation}\label{equa33}
\begin{split}
&{P(A_i)}  = \Pr (\mathop {\max}\limits_{{A_k} \in {\cal{A}}  - \left\{ {{A_i}} \right\}} {{{{| {{h_{{A_k}m}}} |}^2}}} < {{{{| {{h_{{A_i}m}}} |}^2}}})\\
& = 1 + {\sum\limits_{j = 1}^{{2^{| {\cal{A}}  | - 1}} - 1} {\left( { - 1} \right)} ^{| {B(j)} |}} {(1 + {{\sigma _{{A_i}m}^2}}\sum\limits_{{A_k} \in B(j)} {\frac{1}{{\sigma _{{A_k}m}^2}}} )^{ - 1}},
\end{split}
\end{equation}
where ${B(j)}$ represents the $j$-th non-empty subset of $``{\cal{A}}  - \left\{ A_i \right\}"$ and `$-$' represents the set difference.

Moreover, combining (8) and (25), an intercept probability of the MBS-E transmission is obtained as
\begin{equation}\label{equa34}
{P_{{\mathop{M\textrm{int}}} }^{{\textrm{IL}}} = \Pr( {C_{Me}^{{\textrm{IL}}} > R_M^o - R_M^s}),}
\end{equation}
where $R_M^s$ is a secrecy rate of the macro-cell transmission. Substituting $C_{Me}^{{\textrm{IL}}}$ from (8) into (34) yields
\begin{equation}\label{equa35}
{\begin{split}
P_{M{\textrm{int}} }^{{\textrm{IL}}}
& = \Pr (\frac{{{{| {{h_{Ae}}} |}^2}}}{{{\gamma _S}{{| {{h_{Se}}} |}^2} + 1}} > {\Lambda _M})\\
& = \sum\limits_{{A_i} \in {\cal{A}} }^{} {\Pr \left( \begin{split}
&\frac{{{{| {{h_{{A_i}e}}} |}^2}}}{{{\gamma _S}{{| {{h_{Se}}} |}^2} + 1}} > {\Lambda _M},\\
&\mathop {\max}\limits_{{A_k} \in {\cal{A}}  - \left\{ {{A_i}} \right\}} {{{{| {{h_{{A_k}m}}} |}^2}}} < {{{{| {{h_{{A_i}m}}} |}^2}}}
\end{split} \right),}
\end{split}}
\end{equation}
where $\Lambda_M  = ( {{2^{R_M^o - R_M^s}} - 1} )/{\gamma _M}$. Noting that all the random variables ${| {{h_{A_ie}}} |^2}$, ${| {{h_{Se}}} |^2}$ and ${| {{h_{A_im}}} |^2}$ of (35) are independent exponentially distributed random variables with respective means of $\sigma _{A_ie}^2$, $\sigma _{Se}^2$ and $\sigma _{A_im}^2$, we can obtain $P_{M{\textrm{int}} }^{{\textrm{IL}}}$ as
\begin{equation}\label{equa36}
P_{M{\textrm{int}} }^{{\textrm{IL}}} = \sum\limits_{{A_i} \in {\cal{A}} } {P(A_i){P_{M{\textrm{int}},A_i}^{{\textrm{IL}}}}},
\end{equation}
where $P(A_i)$ is given by (33) and ${P_{M{\textrm{int}},A_i}^{{\textrm{IL}}}}$ can be readily computed as
\begin{equation}\label{equa37}
\begin{split}
{P_{M{\textrm{int}},A_i}^{{\textrm{IL}}}}& =  {\Pr (\frac{{{{| {{h_{{A_i}e}}} |}^2}}}{{{\gamma _S}{{| {{h_{Se}}} |}^2} + 1}} > {\Lambda _M})}\\
&=\frac{{\sigma _{{A_i}e}^2}}{{\sigma _{{A_i}e}^2 + \Lambda_M {\gamma _S}\sigma _{Se}^2}}\exp ( - \frac{{{\Lambda _M}}}{{\sigma _{{A_i}e}^2}}).
\end{split}
\end{equation}

Similarly, combining (9) and (25), an intercept probability of the SBS-E transmission is given by
\begin{equation}\label{equa38}
{P_{{\mathop{S\textrm{int}}} }^{{\textrm{IL}}} = \Pr( {C_{Se}^{{\textrm{IL}}} > R_S^o - R_S^s}),}
\end{equation}
where $R_S^s$ is a secrecy rate of the small-cell transmission. Substituting $C_{Se}^{{\textrm{IL}}}$ from (9) into (38) yields
\begin{equation}\label{equa39}
{\begin{split}
P_{S{{\textrm{int}}} }^{{\textrm{IL}}}
& = \Pr (\frac{{{{| {{h_{Se}}} |}^2}}}{{{\gamma _M}{{| {{h_{Ae}}} |}^2} + 1}} > {\Lambda _S})\\
& = \sum\limits_{{A_i} \in {\cal{A}} } {\Pr \left( \begin{split}
&\frac{{{{| {{h_{Se}}} |}^2}}}{{{\gamma _M}{{| {{h_{{A_i}e}}} |}^2} + 1}} > {\Lambda _S},\\
&\mathop {\max}\limits_{{A_k} \in {\cal{A}}  - \left\{ {{A_i}} \right\}} {{{{| {{h_{{A_k}m}}} |}^2}}} < {{{{| {{h_{{A_i}m}}} |}^2}}}
\end{split} \right)} \\
& = \sum\limits_{{A_i} \in {\cal{A}} } P(A_i){P_{S{\textrm{int}} ,A_i}^{{\textrm{IL}}}},
\end{split}}
\end{equation}
where $P(A_i)$ is given by (33) and the term ${P_{S{\textrm{int}} ,A_i}^{{\textrm{IL}}}}$ is obtained as
\begin{equation}\label{equa40}
\begin{split}
{P_{S{\textrm{int}} ,A_i}^{{\textrm{IL}}}} &= {\Pr (\frac{{{{| {{h_{Se}}} |}^2}}}{{{\gamma _M}{{| {{h_{{A_i}e}}} |}^2} + 1}} > {\Lambda _S})}\\
&=\frac{{\sigma _{Se}^2}}{{\sigma _{Se}^2 + {\gamma _M}\sigma _{{A_i}e}^2{\Lambda _S}}}\exp ( - \frac{{{\Lambda _S}}}{{\sigma _{Se}^2}}),
\end{split}
\end{equation}
wherein $\Lambda_S  = ( {{2^{R_S^o - R_S^s}} - 1} )/{\gamma _S}$.

\subsection{Proposed IC-OAS}
This subsection presents the outage probability and intercept probability analysis of macro-cell and small-cell transmissions for the proposed IC-OAS scheme. From (18) and (24), an outage probability of the MBS-MU transmission relying on our IC-OAS scheme is given by
\begin{equation}\label{equa41}
{P_{M\textrm{out}}^{{\textrm{IC}}} = \Pr ( {C_{{Mm}}^{{\textrm{IC}}} < R_M^o} ).}
\end{equation}
Substituting $C_{{Mm}}^{{\textrm{IC}}}$ from (18) into (41) yields
\begin{equation}\label{equa42}
{\begin{split}
P_{M{\textrm{out}}}^{{\textrm{IC}}}& = \Pr [\mathop {\max }\limits_{{A_i} \in {\cal{A}} } \frac{{\sigma _{{A_i}m}^2{\gamma _M} - \sigma _{Sm}^2{\gamma _S}}}{{\sigma _{{A_i}m}^2{\gamma _M}}}{| {{h_{{A_i}m}}} |^2} < {\Delta _M}]\\
& = \prod\limits_{{A_i} \in {\cal{A}} }^{} {\Pr [\frac{{\sigma _{{A_i}m}^2{\gamma _M} - \sigma _{Sm}^2{\gamma _S}}}{{\sigma _{{A_i}m}^2{\gamma _M}}}{{| {{h_{{A_i}m}}} |}^2} < {\Delta _M})]} \\
& = \prod\limits_{{A_i} \in {\cal{A}} }^{} {[1 - \exp ( - \frac{{{\Delta _M}{\gamma _M}}}{{\sigma _{{A_i}m}^2{\gamma _M} - \sigma _{Sm}^2{\gamma _S}}})],}
\end{split}}
\end{equation}
where $\Delta_M  = ( {{2^{R_M^o}} - 1})/{\gamma _M}$. Similarly, by using (20) and (24), an outage probability of the SBS-SU transmission for IC-OAS scheme is expressed as
\begin{equation}\label{equa43}
{P_{S\textrm{out}}^{{\textrm{IC}}} = \Pr ( {C_{{Ss}}^{{\textrm{IC}}} < R_S^o} ),}
\end{equation}
where $R_S^o$ is an overall data rate of the SBS-SU transmission. Substituting $C_{{Ss}}^{{\textrm{IC}}}$ from (20) into (43) and denoting $\Delta_S  = ( {{2^{R_S^o}} - 1})/{\gamma _S}$, we have
{{\begin{equation}\label{equa44}
P_{S{\textrm{out}}}^{{\textrm{IC}}} = \Pr [\frac{{|{h_{Ss}}{|^2}|{h_{Am}}{|^2}}}{{\sigma _{Am}^2}} < ({\Delta _S}{\gamma _M} + \frac{{{X_{Sm}}}}{{\sigma _{Am}^2}})|{h_{As}}{|^2} + {\Delta _S}],
\end{equation}}}
where ${X_{Sm}} = ({2^{R_S^o}} - 1)(|{h_{Sm}}{|^2} - \sigma _{Sm}^2)$. It is very challenging to obtain an exact closed-form expression of $P_{S{\textrm{out}}}^{{\textrm{IC}}}$. Following the existing literature on multi-antenna systems [48]-[50], we assume that the channel fading coefficients $|h_{A_im}|^2$ for different distributed antennas are i.i.d. with the same mean of $\sigma _{{A}m}^2$. Also, the fading coefficients of $|h_{A_is}|^2$ are assumed to be i.i.d. for different distributed antennas, leading to the fact that $|h_{As}|^2$ of (44) follows an exponentially distributed random variable with a mean of $\sigma _{{A}s}^2$, regardless of the selected antenna $A_i$. Moreover, we consider an asymptotic case of ${2^{R_S^o}}\sigma _{Sm}^2 \to 0$, for which the equality of $X_{Sm} = 0$ holds with the probability of one, since both the mean and variance of random variable $X_{Sm}$ approach to zero for ${2^{R_S^o}}\sigma _{Sm}^2 \to 0$. Hence, using (17) and considering the i.i.d. case, we can rewrite (44) as
\begin{equation}\label{equa45}
P^{{\textrm{IC}}}_{S{\textrm{out}}} = \Pr (\frac{{|{h_{Ss}}{|^2}\mathop {\max }\limits_{{A_i} \in {\cal{A}}} |{h_{{A_i}m}}{|^2}}}{{\sigma _{Am}^2}} < {\Delta _S}{\gamma _M}{{|{h_{{A}s}}{|^2}}} + {\Delta _S}),
\end{equation}
for ${2^{R_S^o}}\sigma _{Sm}^2 \to 0$. Letting ${2^{R_S^o}}\sigma _{Sm}^2 \to 0$ and using Appendix B, we obtain $P^{{\textrm{IC}}}_{S{\textrm{out}}}$ from (45) as
\begin{equation}\label{equa46}
\begin{split}
P^{{\textrm{IC}}}_{S{\textrm{out}}} = 1 - &\int_0^\infty  \frac{{N\sigma _{Ss}^2x}}{{\sigma _{Ss}^2x + {\Delta _S}{\gamma _M}\sigma _{As}^2}}\exp ( - \frac{{{\Delta _S}}}{{\sigma _{Ss}^2x}} - x)\\
&\quad\quad\times{{[1 - \exp ( - x)]}^{N - 1}}dx,
\end{split}
\end{equation}
where $N$ is the number of distributed antennas. Moreover, denoting $\beta = \gamma_s/\gamma_M$ and using (B.9) of Appendix B, we can obtain an asymptotic outage probability of $P^{{\textrm{IC}}}_{S{\textrm{out}}} $ in the high SNR region as
\begin{equation}\label{equa47}
\begin{split}
P^{{\textrm{IC}}}_{S{\textrm{out}}} = 1 - &\sum\limits_{k = 0}^{N - 1} \frac{{{{( - 1)}^k}N!}}{{(k + 1)!(N - k - 1)!}}\\
&\quad\quad\times[1 - \Phi _{Ss}^k\exp (\Phi _{Ss}^k)Ei(\Phi _{Ss}^k)],
\end{split}
\end{equation}
for $\gamma_s \to \infty$, wherein $\Phi _{Ss}^k = (k + 1)({2^{R_S^o}} - 1)\sigma _{As}^2/(\beta \sigma _{Ss}^2)$. In addition, combining (22) and (25), an intercept probability of the MBS-E transmission for IC-OAS scheme is obtained as
\begin{equation}\label{equa48}
{P_{{\mathop{M\textrm{int}}} }^{{\textrm{IC}}} = \Pr( {C_{Me}^{{\textrm{IC}}} > R_M^o - R_M^s}).}
\end{equation}
Substituting $C_{Me}^{{\textrm{IC}}}$ from (22) into (48) yields
\begin{equation}\label{equa49}
{\begin{split}
P_{M{\textrm{int}} }^{{\textrm{IC}}} = \Pr \left( {{Z_{Sm}}|{h_{Ae}}{|^2} > {\Lambda _M}({\gamma _S}|{h_{Se}}{|^2}\frac{{|{h_{Am}}{|^2}}}{{\sigma _{Am}^2}} + 1)} \right),
\end{split}}
\end{equation}
where ${Z_{Sm}} = 1 - \frac{\beta }{{\sigma _{Am}^2}}[({2^{R_M^o - R_M^s}} - 1)|{h_{Sm}}{|^2} + \sigma _{Sm}^2]$ and $\Lambda_M  = ( {{2^{R_M^o - R_M^s}} - 1} )/{\gamma _M}$. Assuming that the fading coefficients of $|h_{A_ie}|^2$ are i.i.d. exponentially distributed random variables for different distributed antennas, we can obtain that $|h_{Ae}|^2$ of (49) is exponentially distributed with a mean of $\sigma _{{A}e}^2$. Thus, combining (17) and (49) yields
\begin{equation}\label{equa50}
P_{M{\textrm{int}} }^{{\textrm{IC}}} = \Pr \left( {{Z_{Sm}}|{h_{Ae}}{|^2} > {\Lambda _M}({\gamma _S}|{h_{Se}}{|^2}X + 1)} \right),
\end{equation}
where $X = \frac{{\mathop {\max }\limits_{{A_i} \in {\cal{A}}} |{h_{{A_i}m}}{|^2}}}{{\sigma _{Am}^2}}$. Similarly to (45), we also consider an asymptotic case of ${2^{R_M^o}}\sigma _{Sm}^2 \to 0$, for which the random variable of $({2^{R_M^o - R_M^s}} - 1)|{h_{Sm}}{|^2}$ approaches to $({2^{R_M^o - R_M^s}} - 1)\sigma _{Sm}^2$ with the probability of one, leading to ${Z_{Sm}} = 1 - \frac{\beta }{{\sigma _{Am}^2}}\sigma _{Sm}^2{2^{R_M^o - R_M^s}}$. Noting that ${|{h_{Ae}}{|^2}}$ and ${|{h_{Se}}{|^2}}$ are independent exponentially distributed random variables with respective means of $\sigma^2_{Ae}$ and $\sigma^2_{Se}$, we arrive at
\begin{equation}\label{equa51}
\begin{split}
P_{M{\textrm{int}} }^{{\textrm{IC}}}   = &\int_0^\infty  \frac{{{\Omega _{Sm}}\sigma _{Ae}^2}}{{{\Omega _{Sm}}\sigma _{Ae}^2 + {\Lambda _M}{\gamma _S}\sigma _{Se}^2x}}\\
&\quad\quad\times\exp ( - \frac{{{\Lambda _M}}}{{{\Omega _{Sm}}\sigma _{Ae}^2}}){p_X}(x)dx,
\end{split}
\end{equation}
where ${\Omega _{Sm}} = 1 - \frac{\beta }{{\sigma _{Am}^2}}\sigma _{Sm}^2{2^{R_M^o - R_M^s}}$ and ${{p_X}(x)}$ is the probability density function of $X$ as given by (B.3). Substituting ${{p_X}(x)}$ from (B.3) and (B.7) into (51) gives
\begin{equation}\label{equa52}
\begin{split}
P_{M{\textrm{int}} }^{{\textrm{IC}}}  = &\sum\limits_{k = 0}^{N - 1} \frac{{{{( - 1)}^k}(N - 1)!}}{{k!(N - k - 1)!}}\exp ( - \frac{{{\Lambda _M}}}{{{\Omega _{Sm}}\sigma _{Ae}^2}})\\
&\quad\times\int_0^\infty  {\frac{{\Phi _{Ae}^kN}}{{\Phi _{Ae}^k + (k + 1)x}}\exp [ - (k + 1)x]dx} ,
\end{split}
\end{equation}
where $\Phi _{Ae}^k = \frac{{(k + 1){\Omega _{Sm}}\sigma _{Ae}^2}}{{({2^{R_M^o - R_M^s}} - 1)\beta \sigma _{Se}^2}}$. Substituting $x = (t - \Phi _{Ae}^k)/(k + 1)$ into the preceding equation and performing the integration yield
\begin{equation}\label{equa53}
\begin{split}
P_{M{\textrm{int}} }^{{\textrm{IC}}}   = &\sum\limits_{k = 0}^{N - 1} \frac{{{{( - 1)}^k}N!\Phi _{Ae}^k}}{{(k + 1)!(N - k - 1)!}}\\
&\quad\quad\times\exp (\Phi _{Ae}^k - \frac{{{\Lambda _M}}}{{{\Omega _{Sm}}\sigma _{Ae}^2}})Ei(\Phi _{Ae}^k).
\end{split}
\end{equation}
Similarly, combining (23) and (25), we can obtain an intercept probability of the SBS-E transmission as
\begin{equation}\label{equa54}
P_{S{\textrm{int}} }^{{\textrm{IC}}} = \Pr \left( {\frac{{|{h_{Se}}{|^2}|{h_{Am}}{|^2}}}{{\sigma _{Am}^2}} > ({\Lambda _S}{\gamma _M} + \frac{{X_{Sm}^{'}}}{{\sigma _{Am}^2}})|{h_{Ae}}{|^2} + {\Lambda _S}} \right),
\end{equation}
where ${\Lambda _S} = {{{2^{R_S^o - R_S^s}} - 1}}/{{{\gamma _S}}}$ and $X_{Sm}^{'} = ({2^{R_S^o - R_S^s}} - 1)(|{h_{Sm}}{|^2} - \sigma _{Sm}^2)$. It is challenging to obtain a general closed-form expression for $P_{S{\textrm{int}}}^{{\textrm{IC}}}$. Similarly, we assume that the fading coefficients of $|h_{A_ie}|^2$ for different distributed antennas are i.i.d. exponentially distributed, thus $|h_{Ae}|^2$ of (54) is exponentially distributed with a mean of $\sigma _{{A}e}^2$. Moreover, we consider an asymptotic case of ${2^{R_S^o}}\sigma _{Sm}^2 \to 0$, for which an equality of ${X_{Sm}^{'}} = 0$ holds with the probability of one. As a consequence, combining (17) and (54) yields
\begin{equation}\label{equa55}
P_{S{\textrm{int}} }^{{\textrm{IC}}} =\Pr \left( {|{h_{Se}}{|^2}X > {\Lambda _S}{\gamma _M}|{h_{Ae}}{|^2} + {\Lambda _S}} \right),
\end{equation}
for ${2^{R_S^o}}\sigma _{Sm}^2 \to 0$, where $X = \frac{{\mathop {\max }\limits_{{A_i} \in {\cal{A}}} |{h_{{A_i}m}}{|^2}}}{{\sigma _{Am}^2}}$. Noting that ${|{h_{Ae}}{|^2}}$ and ${|{h_{Se}}{|^2}}$ are independent exponentially distributed random variables with respective means of $\sigma^2_{Ae}$ and $\sigma^2_{Se}$, we obtain
\begin{equation}\label{equa56}
P_{S{\textrm{int}} }^{{\textrm{IC}}}  = \int_0^\infty  {\frac{{\sigma _{Se}^2x}}{{\sigma _{Se}^2x + {\Lambda _S}{\gamma _M}\sigma _{Ae}^2}}\exp ( - \frac{{{\Lambda _S}}}{{\sigma _{Se}^2x}}){p_X}(x)dx},
\end{equation}
where ${{p_X}(x)}$ is the probability density function of $X$ as given by (B.3). Substituting ${{p_X}(x)}$ from (B.3) and (B.7) into (56) and letting $\gamma_S \to \infty$, we can simplify (56) as
\begin{equation}\label{equa57}
\begin{split}
P_{S{\textrm{int}} }^{{\textrm{IC}}} =& \sum\limits_{k = 0}^{N - 1} \frac{{{{( - 1)}^k}N!}}{{k!(N - k - 1)!}}\int_0^\infty  \frac{{(k + 1)x}}{{(k + 1)x + \Phi _{Se}^k}}\\
&\quad\quad\quad\quad\quad\quad\quad\quad\quad\quad\times\exp [ - (k + 1)x]dx ,
\end{split}
\end{equation}
where $\Phi _{Se}^k = (k + 1)({2^{R_S^o - R_S^s}} - 1)\sigma _{Ae}^2/(\beta \sigma _{Se}^2)$. Performing the integral of (57) yields
\begin{equation}\label{equa58}
P_{S{\textrm{int}} }^{{\textrm{IC}}}  = \sum\limits_{k = 0}^{N - 1} {\frac{{{{( - 1)}^k}N![1 - \Phi _{Se}^k\exp (\Phi _{Se}^k)Ei(\Phi _{Se}^k)]}}{{(k + 1)!(N - k - 1)!}}},
\end{equation}
where $Ei(x) = \int_x^\infty  {\frac{1}{t}\exp ( - t)dt} $ is known as the exponential integral function.

\section{Numerical Results And Discussions}
In this section, we present numerical SRT results of IL-OAS and IC-OAS schemes in terms of their outage probability and intercept probability for both the macro cell and small cell. For national convenience, let $\beta  = {P_s}/{P_M}$ denote a ratio of the transmit power of SBS to that of MBS, called SMR for short. In our numerical evaluation, fading variances of the main channel, interference channel and wiretap channel are given by one, i.e., $E(|{g_{{A_i}m}}{|^2}) = E(|{g_{{A_i}s}}{|^2}) = E(|{g_{{A_i}e}}{|^2}) = E(|{g_{Ss}}{|^2}) =  E(|{g_{Sm}}{|^2}) =  E(|{g_{Se}}{|^2}) = 1$. The transmission distances of ${d_{{A_i}m}} = {d_{{A_i}s}} = {d_{{A_i}e}}  = {d_{Sm}} = {d_{Se}} = 300{{m}}$ are assumed, unless otherwise stated. Since a small cell typically has a much narrower coverage than a macro cell, a distance of ${d_{Ss}} = 30{m}$ is used for the SBS-SU transmission. Moreover, path loss factors of $\alpha_{{A_i}m} = \alpha_{Ss} = \alpha_{{A_i}e} = \alpha_{Se} = 2.5$ are assumed, while $\alpha_{{A_i}s} = \alpha_{Sm} = 3.5$ are specified for the cross-interference channels between the macro cell and small cell, considering that the small cell is deployed in a shadowed area (e.g., in-building area, underground garage, etc.) of the macro cell. Additionally, the number of distributed antennas of $N = 16$, an SNR of $\gamma_M = 70$dB, a secrecy data rata of $R_M^s =R_S^s = 1{\textrm{bit/s/Hz}}$, and an SMR of $\beta = 0.1$ are assumed, unless otherwise mentioned. It is pointed out that both theoretical and simulated SRT results are given in the following Figs. 2-8, where the theoretical outage probabilities and intercept probabilities of IL-OAS and IC-OAS schemes are computed by using (28), (31), (36), (39), (42), (46), (53), and (56), respectively, and the corresponding simulated results are obtained through Monte-Carlo simulations. As observed from Figs. 2-8, the theoretical and simulated results match well in terms of the outage probability and intercept probability, validating the correctness of our theoretical SRT analysis.

\begin{figure}
\centering
\includegraphics[scale = 0.55]{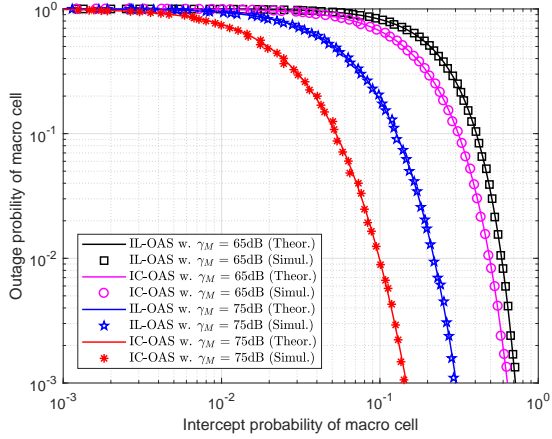}
\caption{{{Outage probability versus intercept probability of the macro cell with the conventional IL-OAS and proposed IC-OAS schemes with $R_M^s=R_S^s=1$bit/s/Hz and $R_M^o=R_S^o$ $\in$ $[ {R_M^s,10}]$ for different SNRs of $\gamma_M = 65{\textrm{dB}}$ and $75\textrm{dB}$.}}}
\label{fig 2}
\end{figure}

In Fig. 2, we show the outage probability versus intercept probability of the macro cell with the conventional IL-OAS and proposed IC-OAS schemes for different SNRs of $\gamma_M = 65{\textrm{dB}}$ and $75\textrm{dB}$. It can be seen from Fig. 2 that as the intercept probability increases, outage probabilities of both the IL-OAS and IC-OAS schemes decrease, and vice versa. In other words, the transmission reliability can be improved at the cost of a security degradation, meaning a tradeoff between the security and reliability, referred to as the security-reliability tradeoff (SRT). Fig. 2 also shows that for both cases of $\gamma_M = 65$dB and $75$dB, the proposed IC-OAS scheme outperforms the conventional IL-OAS method in terms of the SRT of macro cell. Moreover, as the SNR $\gamma_M$ increases from $65$dB to $75$dB, the SRT performance advantage of proposed IC-OAS scheme over conventional IL-OAS becomes more significant.

\begin{figure}
\centering
\includegraphics[scale = 0.55]{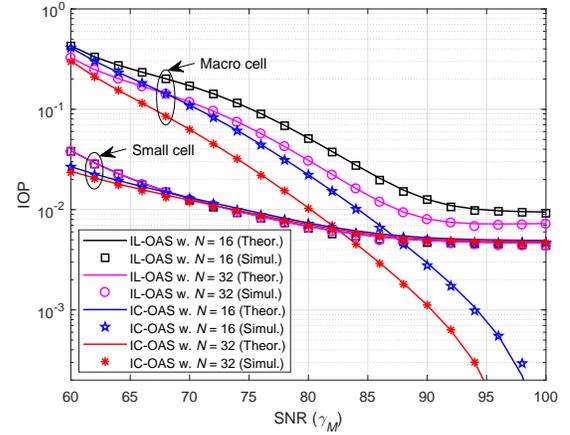}
{\caption{IOP versus SNR $\gamma_M$ of the macro cell and small cell with the IL-OAS and IC-OAS schemes for different number of distributed antennas of $N=16$ and $32$, where the IOP stands for the mean of intercept probability and outage probability.}}
\label{fig 3}
\end{figure}

Fig. 3 shows the mean of intercept probability and outage probability (denoted by IOP for short) versus SNR $\gamma_M$ of the macro cell and small cell with the IL-OAS and IC-OAS schemes for different number of distributed antennas of $N=16$ and $32$. It is noted that given an SNR of $\gamma_M$, the numerical IOP results of IL-OAS and IC-OAS schemes are minimized for the macro cell and small cell through adjusting overall data rates of $R^o_M$ and $R^o_S$, respectively. One can observe from Fig. 3 that for both cases of $N=16$ and $32$, the proposed IC-OAS scheme significantly outperforms the conventional IL-OAS method in terms of the IOP of macro cell. Moreover, as the SNR $\gamma_M$ increases, the IOP of conventional IL-OAS scheme gradually decreases to a floor value, whereas the proposed IC-OAS scheme continuously improves the IOP of macro cell without the floor effect. This implies that the SRT performance of macro cell relying on our IC-OAS scheme can be improved by simply increasing the transmit power of $P_M$. In addition, it is also seen from Fig. 3 that as the SNR increases, the IOP of small cell with IC-OAS is initially better than that with IL-OAS, which eventually converge toward each other in the high SNR region. Hence, as compared to the conventional IL-OAS method, the proposed IC-OAS scheme not only brings SRT benefits to the macro cell, but also improves the SRT of small cell in the low SNR region.

\begin{figure}
\centering
\includegraphics[scale = 0.55]{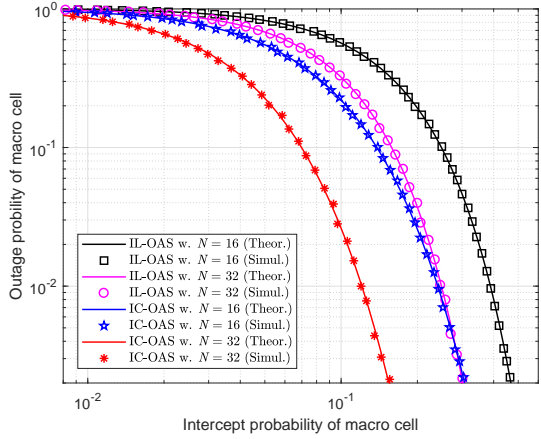}
{\caption{Outage probability versus intercept probability of the macro cell with the IL-OAS and IC-OAS schemes for different number of distributed antennas of $N=16$ and $32$.}}
\label{fig 4}
\end{figure}
Fig. 4 depicts the outage probability versus intercept probability of macro cell with the IL-OAS and IC-OAS schemes for different number of distributed antennas of $N=16$ and $32$. It is shown from Fig. 4 that for both cases of $N=16$ and $32$, the SRT performance of proposed IC-OAS scheme is always better than that of conventional IL-OAS method. Moreover, as the number of distributed antennas increases from $N=16$ to $32$, the SRT gap between the IL-OAS and IC-OAS schemes enlarges, meaning more SRT improvement achieved by the proposed IC-OAS with an increasing number of distributed antennas, compared with the conventional IL-OAS.

\begin{figure}
\centering
\includegraphics[scale = 0.55]{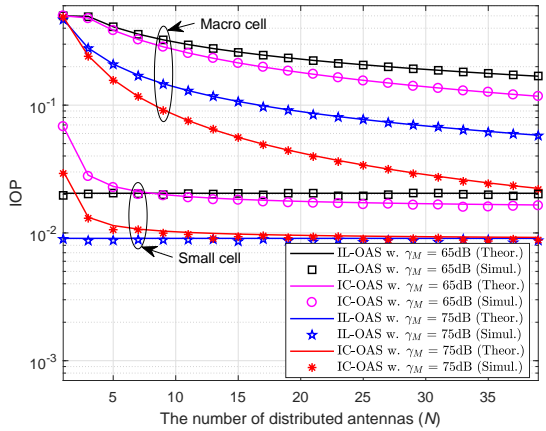}
{\caption{IOP versus the number of distributed antennas $N$ of the macro cell and small cell with the IL-OAS and IC-OAS schemes for different SNRs of $\gamma_M = 65{\textrm{dB}}$ and $75\textrm{dB}$, where the IOP stands for the mean of intercept probability and outage probability.}}
\end{figure}

In order to further demonstrate the impact of the number of distributed antennas $N$ on the intercept and outage probability, Fig. 5 shows IOPs of the macro cell and small cell versus the number of distributed antennas $N$ for the IL-OAS and IC-OAS schemes. One can observe from Fig. 5 that for both cases of $\gamma_M = 65{\textrm{dB}}$ and $75\textrm{dB}$, as the number of distributed antennas increases, the IOPs of macro cell with both IL-OAS and IC-OAS schemes are improved and the performance advantage of IC-OAS over IL-OAS increases accordingly. Moreover, it can be seen from Fig. 5 that the IOP of small cell for the conventional IL-OAS method keeps unchanged and has no improvement, as the number of distributed antennas increases. By contrast, the proposed IC-OAS scheme can decrease the IOP of small cell with an increasing number of distributed antennas, which even has a better IOP performance than the IL-OAS for $N \ge 10$ in the case of $\gamma_M = 65{\textrm{dB}}$, as shown from Fig. 5. As a consequence, one can conclude from Figs. 3 and 5 that compared with the conventional IL-OAS, the proposed IC-OAS scheme not only improves the SRT of macro cell, but also enhances the SRT of small cell in the low SNR region through increasing the number of distributed antennas.

\begin{figure}
\centering
\includegraphics[scale = 0.55]{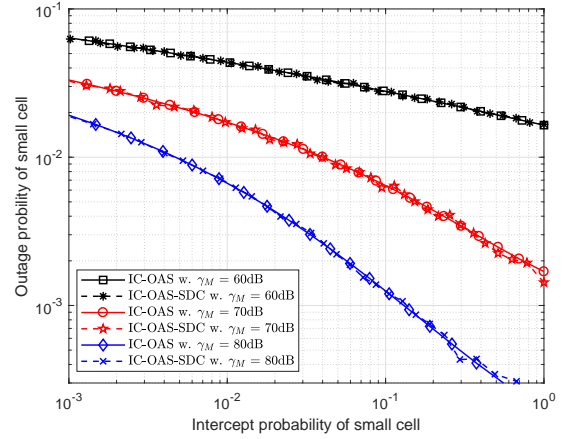}
{\caption{Outage probability versus intercept probability of the small cell with the IC-OAS and IC-OAS-SDC schemes for different SNRs of $\gamma_M = 60{\textrm{dB}}$, $ 70{\textrm{dB}}$, and $80\textrm{dB}$.}}
\end{figure}
Considering that the designed signal $x_{A_i}$ of (12) contains the SBS' signal $x_S$, we are intended to examine the SRT of small cell for the case that the SU and eavesdropper both leverage an additional information contained in $x_{A_i}$ to decode the SBS' signal. To be specific, the SU employs the selection diversity combining (SDC) to jointly exploit the terms ${h_{As}}x_{A}$ and ${h_{Ss}}\sqrt {{P_S}} w_S {x_S}$ of (19) for decoding $x_S$, where either ${h_{As}}x_{A}$ or ${h_{Ss}}\sqrt {{P_S}} w_S {x_S}$ is opportunistically utilized depending on which has a higher SNR. Also, the eavesdropper is considered to adopt a similar SDC method in leveraging ${h_{Ae}}x_{A}$ and ${h_{Se}}\sqrt {{P_S}} w_S {x_S}$ of (21) for tapping the SBS' signal $x_S$. The combination of the aforementioned SDC process with IC-OAS is denoted by IC-OAS-SDC for short. Fig. 6 shows the outage probability versus intercept probability of the small cell with the IC-OAS and IC-OAS-SDC schemes for different SNRs of $\gamma_M = 60{\textrm{dB}}$, $ 70{\textrm{dB}}$, and $80\textrm{dB}$, where the SRT results of IC-OAS-SDC are obtained through Monte-Carlo simulations. It is illustrated from Fig. 6 that the IC-OAS-SDC scheme achieves the same performance as the IC-OAS without any SRT benefits for all the case of $\gamma_M = 60{\textrm{dB}}$, $ 70{\textrm{dB}}$, and $80\textrm{dB}$. This is due to the fact that although the SDC is employed at the SU to extract the SBS' signal from the designed signal $x_{A_i}$ for improving the transmission reliability of small cell, it can be similarly adopted by the eavesdropper for degrading the secrecy, thus no extra SRT improvement is expected for the small cell.

\begin{figure}
\centering
\includegraphics[scale = 0.55]{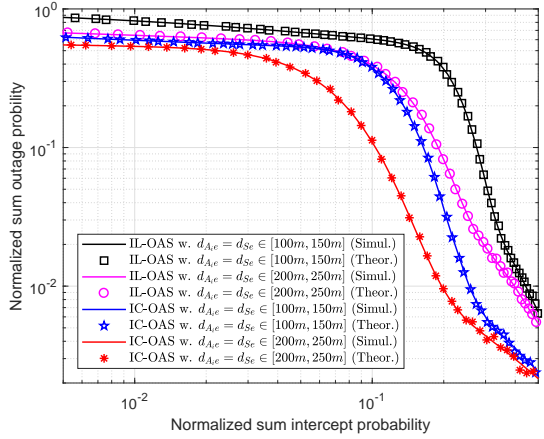}
{\caption{Normalized sum outage probability versus normalized sum intercept probability of the macro cell and small cell for the IL-OAS and IC-OAS schemes with $R_M^s=R_S^s=1$bit/s/Hz and $R_M^o=R_S^o$ $\in$ $[ {R_M^s,10}]$ for different independent uniformly distributed distances of $d_{A_ie}$ and $d_{Se}$, where the normalized sum outage probability is the mean of individual outage probabilities of the macro cell and small cell and the normalized sum intercept probability is the mean of their individual intercept probabilities.}
\label{fig 7}}
\end{figure}
Although Figs. 2-6 demonstrate that the proposed IC-OAS scheme is capable of improving the SRTs of both the macro cell and small cell as compared to the conventional IL-OAS method, they do not provide an overall SRT performance of the heterogeneous cellular network by taking into account the macro cell and small cell jointly. To this end, we show a normalized sum outage probability versus sum intercept probability of the macro cell and small cell for the IL-OAS and IC-OAS schemes in Fig. 7. To be specific, the normalized sum outage probability is defined as an average value of individual outage probabilities of the macro cell and small cell, while a mean of their individual intercept probabilities is considered as the normalize sum intercept probability. Since the eavesdropper may randomly move around with an unknown position, we here consider that the transmission distances of ${A_i}$-E and SBS-E (i.e., $d_{A_ie}$ and $d_{Se}$) are independent uniformly distributed. As seen from Fig. 7, given a sum intercept probability requirement, the sum outage probability of proposed IC-OAS scheme is lower than that of IL-OAS method and vice versa, showing an overall SRT improvement for the heterogeneous cellular network. Moreover, as the eavesdropper's moving range increases from $d_{A_ie}=d_{Se} \in [100m,150m]$ to $[200m,250m]$, the overall SRTs of IL-OAS and IC-OAS schemes are improved slightly. This is due to the fact that when the eavesdropper moves away from MBS and SBS, it has a worsened signal reception quality along with an enhanced secrecy performance, thus an improved overall SRT is achieved for the heterogeneous cellular network.

\begin{figure}
\centering
\includegraphics[scale = 0.55]{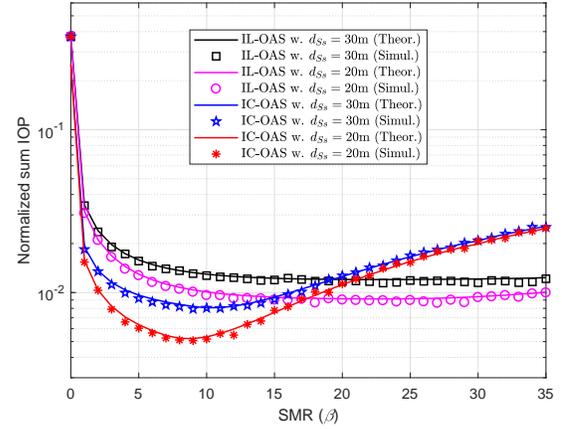}
{\caption{Normalized sum IOP versus SMR of the macro cell and small cell with the IL-OAS and IC-OAS schemes for different SBS-SU transmission distances of $d_{Ss}=20m$ and $30m$, where the normalized sum IOP is the mean of the individual IOPs of the macro cell and small cell.}
\label{fig8}}
\end{figure}
Fig. 8 shows the normalized sum IOP versus SMR of the macro cell and small cell with the IL-OAS and IC-OAS schemes for different SBS-SU transmission distances of $d_{Ss}=20m$ and $30m$. It needs to be pointed out that as implied from (14), the SMR $\beta$ should satisfy an inequality of $\beta  \le \sigma _{A_im}^2/\sigma _{Sm}^2$ for completely canceling out the mutual interference received at MU. Noting that the fading variances of $E(|{g_{{A_i}m}}{|^2})= E(|{g_{Sm}}{|^2}) = 1$, the transmission distances of ${d_{{A_i}m}} = {d_{Sm}} = 300{{m}}$, the path loss factors of ${\alpha_{{A_i}m}}= 2.5$ and $ {\alpha_{Sm}} = 3.5$ are considered in our numerical evaluation, we can readily obtain that the SMR should be in a range of $0 \le \beta  \le 300$. As shown in Fig. 8, for both cases of $d_{Ss}=20m$ and $30m$, as the SMR increases, the sum IOP of IL-OAS initially decreases and then remains almost constant. This is because that with an increasing SMR, a higher transmit power is used at SBS, which leads to the fact that the SRT performance of small cell gradually improves to an SRT floor. In regard to the macro cell, as the transmit power of SBS increases, more interference is encountered at the eavesdropper in tapping the MBS-MU transmission and thus the secrecy performance of macro cell is enhanced. Meanwhile, with an increasing transmit power of SBS, MU would also receive more interference from SBS, resulting in an outage performance degradation. Overall speaking, in the conventional IL-OAS approach, the secrecy improvement would be mostly neutralized with the outage degradation for the macro cell. Hence, by jointly considering the macro cell and small cell, an overall SRT of the IL-OAS scheme gradually converges in the high SMR region.

By contrast, as the SMR increases, the outage degradation of macro cell with our IC-OAS scheme is alleviated due to the adopted interference cancelation mechanism which neutralizes an increased interference received at MU from SBS, leading to an initial decrease of the sum IOP for the IC-OAS scheme. However, it comes at the cost of MBS' power resources, since partial transmit power of MBS is consumed to emit a specially-designed signal $x_{A_i}$ for the interference cancelation. As an extreme case, when the SMR increases to $\beta = \sigma _{A_im}^2/\sigma _{Sm}^2$ (i.e., $P_s = P_M$), all the transmit power of MBS is allocated for emitting $x_{A_i}$ to cancel out the interference received at MU as implied from (13), and no transmit power is left for sending the information-bearing signal of $x_M$, resulting in an outage probability of one for the macro cell. It can be concluded that as the SMR continues to increase after a sufficiently high value, the outage degradation dominates the secrecy enhancement for the macro cell in our IC-OAS scheme. Consequently, the overall SRT performance of heterogeneous cellular networks relying on our IC-OAS scheme can be further optimized with regard to the SMR in terms of minimizing the normalized sum IOP. Additionally, as shown from Fig. 8, for both cases of $d_{Ss}=20m$ and $30m$, the optimized IOP performance of proposed IC-OAS scheme is always much better than that of conventional IL-OAS method.

\section{Conclusion}
In this paper, we investigated physical-layer security for a heterogeneous cellular network, where a macro cell coexists with a small cell in the face of a common passive eavesdropper. We proposed an interference-canceled opportunistic antenna selection (IC-OAS) scheme to enhance physical-layer security for the aforementioned heterogeneous cellular network. Meanwhile, the conventional interference-limited OAS (IL-OAS) approach was considered as a benchmark. Specifically, in the proposed IC-OAS scheme, MBS transmits its confidential message to MU with the help of an opportunistic distributed antenna, where a special signal is artificially designed and emitted at MBS to cancel out the interference received at MU from SBS. An SRT analysis was carried out to evaluate the performance of IL-OAS and IC-OAS schemes in terms of the outage probability and intercept probability. Numerical results illustrated that compared with the conventional IL-OAS, the proposed IC-OAS scheme is capable of improving SRTs of both the macro cell and small cell by employing more distributed antennas. Moreover, by jointly taking into account the macro cell and small cell, the proposed IC-OAS scheme significantly outperforms the conventional IL-OAS in terms of the normalized sum IOP. Additionally, it was shown that the normalized sum IOP of IC-OAS can be further optimized with regard to the SMR and the optimized sum IOP of proposed IC-OAS scheme is much better than that of conventional IL-OAS method.

\section*{Appendix A}
\section*{Derivation of (28)}
Without loss of generality, we have the following notations ${{X_{{i}}} = {{{{| {{h_{{A_i}m}}} |}^2}}}}$, ${X = \mathop {\max}\limits_{{A_i} \in {\cal{A}}}{{{{| {{h_{{A_i}m}}} |}^2}}}}$ and $Y = {{{{| {{h_{Sm}}} |}^2}}}$. Using (27) and letting ${f_{{X_{{i}}}}}({x_{{i}}})$, ${f_{{X}}}\left( {{x}} \right)$ and ${f_{{Y}}}\left( {{y}} \right)$ denote the probability density functions (PDFs) of ${X_{{i}}}$, $X$ and $Y$, respectively, we have
\begin{equation}\renewcommand\theequation{A.1}
\begin{split}
P_{M{\textrm{out}}}^{{\textrm{IL}}} &= \Pr (X < {\Delta _M}{\gamma _S}Y + {\Delta _M})\\
&= \int_0^\infty  {{F_X}} ({\Delta _M}{\gamma _S}y + {\Delta _M}){f_Y}(y)dy,
\end{split}
\end{equation}
where ${F_X}(x) = \Pr(X < x)$. Since ${| {{h_{A_im}}} |^2}$ and ${| {{h_{Sm}}} |^2}$ are independent exponentially distributed random variable with respective means of $\sigma _{A_im}^2$ and $\sigma _{Sm}^2$, ${f_{{X_{{i}}}}}({x_{{i}}})$ and ${f_{{Y}}}\left( {{y}} \right)$ can be given by
\begin{equation}\renewcommand\theequation{A.2}
{f_{X_{i}}}\left( x_{i} \right) = \frac{1}{{\sigma _{{A_i}m}^2 }}\exp ( - \frac{x_{i}}{{\sigma _{{A_i}m}^2}}),
\end{equation}
and
\begin{equation}\renewcommand\theequation{A.3}
{f_Y}\left( y \right) = \frac{1}{{\sigma _{Sm}^2 }}\exp ( - \frac{y}{{\sigma _{Sm}^2}}).
\end{equation}
Moreover, using (A.2), we can readily obtain ${F_X}(x)$ as
\begin{equation}\renewcommand\theequation{A.4}
\begin{split}
&{F_X}(x) = \Pr(\mathop {\max }\limits_{{A_i} \in {\cal{A}} } {|h_{{A_i}}|^2} < x)\\
&= 1 + \sum\limits_{j = 1}^{{2^{|{\cal{A}} |}} - 1} {{{( - 1)}^{|A(j)|}}} \exp ( - \sum\limits_{{A_i} \in {\cal{A}}(j)} {\frac{z}{{\sigma _{{A_i}m}^2}}}),
\end{split}
\end{equation}
where ${A(j)}$ represents the $j$-th non-empty subset of the antenna set ${\cal{A}}$. Substituting (A.3) and (A.4) into (A.1) and performing the integration of (A.1) yields $P_{S{\textrm{out}}}^{{\textrm{IL}}}$ of (28).

\section*{Appendix B}
\section*{Derivation of (46) and (47)}
Denoting $X = {{\mathop {\max }\limits_{{A_i} \in {\cal{A}}} |{h_{{A_i}m}}{|^2}}}/{{\sigma _{Am}^2}}$, we can rewrite (45) as
\begin{equation}\renewcommand\theequation{B.1}
{\mathop{\rm P}\nolimits} _{S{\textrm{out}}}^{{\textrm{IC}}} = \Pr (|{h_{Ss}}{|^2}X < {\Delta _S}{\gamma _M}|{h_{As}}{|^2} + {\Delta _S}).
\end{equation}
Since the fading coefficients $|{h_{{A_i}m}}{|^2}$ for different antennas are assumed to be i.i.d. with a mean of $\sigma _{Am}^2$, we obtain the cumulative distribution function of $X$ as
\begin{equation}\renewcommand\theequation{B.2}
\begin{split}
\Pr (X < x) &= \Pr (\mathop {\max }\limits_{{A_i} \in {\cal{A}}} |{h_{{A_i}m}}{|^2} < \sigma _{Am}^2x)\\
& = {[1 - \exp ( - x)]^N},
\end{split}
\end{equation}
where $N$ is the number of distributed antennas. From (B.2), the probability density function of $X$ is given by
\begin{equation}\renewcommand\theequation{B.3}
{p_X}(x) = N\exp ( - x){[1 - \exp ( - x)]^{N - 1}}.
\end{equation}
Noting that $|{h_{Ss}}{|^2}$ and $|{h_{As}}{|^2}$ are independent exponentially distributed random variables with respective means of $\sigma^2_{Ss}$ and $\sigma^2_{As}$, we obtain an outage probability of the SBS-SU transmission for IC-OAS scheme from (B.1) as
\begin{equation}\renewcommand\theequation{B.4}
{\mathop{\rm P}\nolimits} _{S{\textrm{out}}}^{{\textrm{IC}}} = 1 - \int_0^\infty  {\frac{{\sigma _{Ss}^2x}}{{\sigma _{Ss}^2x + {\Delta _S}{\gamma _M}\sigma _{As}^2}}\exp ( - \frac{{{\Delta _S}}}{{\sigma _{Ss}^2x}}){p_X}(x)dx}.
\end{equation}
Substituting ${p_X}(x)$ from (B.3) into (B.4) yields
\begin{equation}\renewcommand\theequation{B.5}
\begin{split}
{\mathop{\rm P}\nolimits}_{S{\textrm{out}}}^{{\textrm{IC}}} = 1 -& \int_0^\infty  \frac{{N\sigma _{Ss}^2x}}{{\sigma _{Ss}^2x + {\Delta _S}{\gamma _M}\sigma _{As}^2}}\exp ( - \frac{{{\Delta _S}}}{{\sigma _{Ss}^2x}} - x)\\
&\quad\quad\times{{[1 - \exp ( - x)]}^{N - 1}}dx,
\end{split}
\end{equation}
which is (46). Moreover, the following presents an asymptotic outage probability analysis of the SBS-SU transmission for IC-OAS scheme in the high SNR region as $(\gamma_M, \gamma_S) \to \infty$, for which $\Delta_S = 0$ holds. Denoting $\beta = \gamma_S/\gamma_M$ and substituting $\Delta_S = 0$ into (B.5) yield
\begin{equation}\renewcommand\theequation{B.6}
\begin{split}
{\mathop{\rm P}\nolimits} _{S{\textrm{out}}}^{{\textrm{IC}}} = 1 - &\int_0^\infty  \frac{{N\sigma _{Ss}^2x}}{{\sigma _{Ss}^2x + ({2^{R_S^o}} - 1)\sigma _{As}^2/\beta }}\exp ( - x)\\
&\quad\quad\times{{[1 - \exp ( - x)]}^{N - 1}}dx,
\end{split}
\end{equation}
for $\gamma_S \to \infty$. Using the binomial theorem, we have
\begin{equation}\renewcommand\theequation{B.7}
{[1 - \exp ( - x)]^{N - 1}} = \sum\limits_{k = 0}^{N - 1} {\frac{{{{( - 1)}^k}(N - 1)!}}{{k!(N - k - 1)!}}\exp ( - kx)}.
\end{equation}
Combining (B.6) and (B.7) gives
\begin{equation}\renewcommand\theequation{B.8}
\begin{split}
{\mathop{\rm P}\nolimits} _{S{\textrm{out}}}^{{\textrm{IC}}} = 1 -& \sum\limits_{k = 0}^{N - 1} \frac{{{{( - 1)}^k}N!}}{{k!(N - k - 1)!}}\\
&\quad\quad\times\int_0^\infty  {\frac{{\sigma _{Ss}^2x\exp [ - (k + 1)x]}}{{\sigma _{Ss}^2x + ({2^{R_S^o}} - 1)\sigma _{As}^2/\beta }}dx} .
\end{split}
\end{equation}
Moreover, denoting $\Phi _{Ss}^k = (k + 1)({2^{R_S^o}} - 1)\sigma _{As}^2/(\beta \sigma _{Ss}^2)$ and substituting $x = (t - \Phi _{Ss}^k)/(k + 1)$ into (B.8), we arrive at
\begin{equation}\renewcommand\theequation{B.9}
\begin{split}
{\mathop{\rm P}\nolimits} _{S{\textrm{out}}}^{{\textrm{IC}}} &= 1 - \sum\limits_{k = 0}^{N - 1} \frac{{{{( - 1)}^k}N!}}{{(k + 1)!(N - k - 1)!}}\\
&\quad\quad\quad\quad\quad\times\int_{\Phi _{Ss}^k}^\infty  {(1 - \frac{{\Phi _{Ss}^k}}{t})\exp (\Phi _{Ss}^k - t)dt}   \\
&= 1 - \sum\limits_{k = 0}^{N - 1} \frac{{{{( - 1)}^k}N!}}{{(k + 1)!(N - k - 1)!}}\\
&\quad\quad\quad\quad\quad\times[1 - \Phi _{Ss}^k\exp (\Phi _{Ss}^k)Ei(\Phi _{Ss}^k)],
\end{split}
\end{equation}
where $Ei(x) = \int_x^\infty  {\frac{1}{t}\exp ( - t)dt} $ is known as the exponential integral function.

\begin{IEEEbiography}[{\includegraphics[width=1in,height=1.25in]{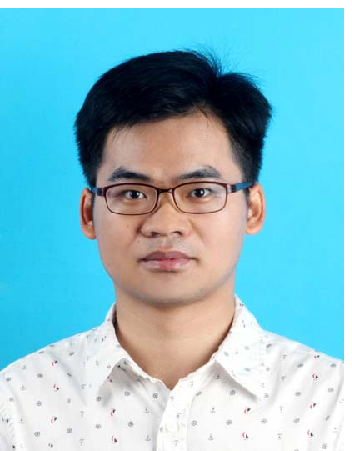}}]{Yulong Zou} (SM'13) is a Full Professor and Doctoral Supervisor at the Nanjing University of Posts and Telecommunications (NUPT), Nanjing, China. He received the B.Eng. degree in information engineering from NUPT, Nanjing, China, in July 2006, the first Ph.D. degree in electrical engineering from the Stevens Institute of Technology, New Jersey, USA, in May 2012, and the second Ph.D. degree in signal and information processing from NUPT, Nanjing, China, in July 2012.

Dr. Zou was awarded the 9th IEEE Communications Society Asia-Pacific Best Young Researcher in 2014. He has served as an editor for the IEEE Communications Surveys \& Tutorials, IEEE Communications Letters, EURASIP Journal on Advances in Signal Processing, IET Communications, and China Communications. In addition, he has acted as TPC members for various IEEE sponsored conferences, e.g., IEEE ICC/GLOBECOM/WCNC/VTC/ICCC, etc.
\end{IEEEbiography}

\begin{IEEEbiography}[{\includegraphics[width=1in,height=1.25in]{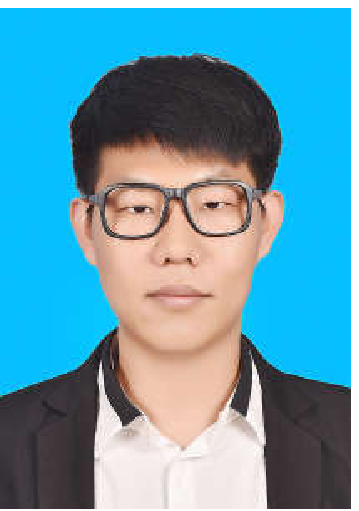}}]{Ming Sun} received the B.S degree with major on Communication Engineering from Nantong University (NTU), Nantong, China, in July 2016. He is currently pursuing the M.S. degree in Signal and Information Processing at the Nanjing University of Posts and Telecommunications (NUPT). His research interests include cognitive radio, cooperative communications, and wireless security.\end{IEEEbiography}

\begin{IEEEbiography}[{\includegraphics[width=1in,height=1.25in]{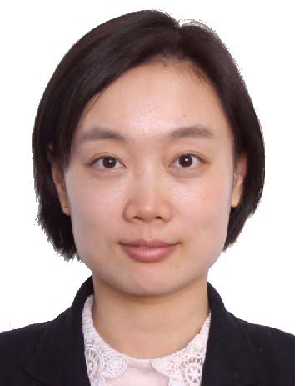}}]{Jia Zhu} is an Associate Professor at the Nanjing University of Posts and Telecommunications (NUPT), Nanjing, China. She received the B.Eng. degree in Computer Science and Technology from the Hohai University, Nanjing, China, in July 2005, and the Ph.D. degree in Signal and Information Processing from the Nanjing University of Posts and Telecommunications, Nanjing, China, in April 2010. From June 2010 to June 2012, she was a Postdoctoral Research Fellow at the Stevens Institute of Technology, New Jersey, the United States. Since November 2012, she has been a full-time faculty member with the Telecommunication and Information School of NUPT, Nanjing, China. Her general research interests include the cognitive radio, physical-layer security and communications theory.\end{IEEEbiography}

\begin{IEEEbiography}[{\includegraphics[width=1in,height=1.25in]{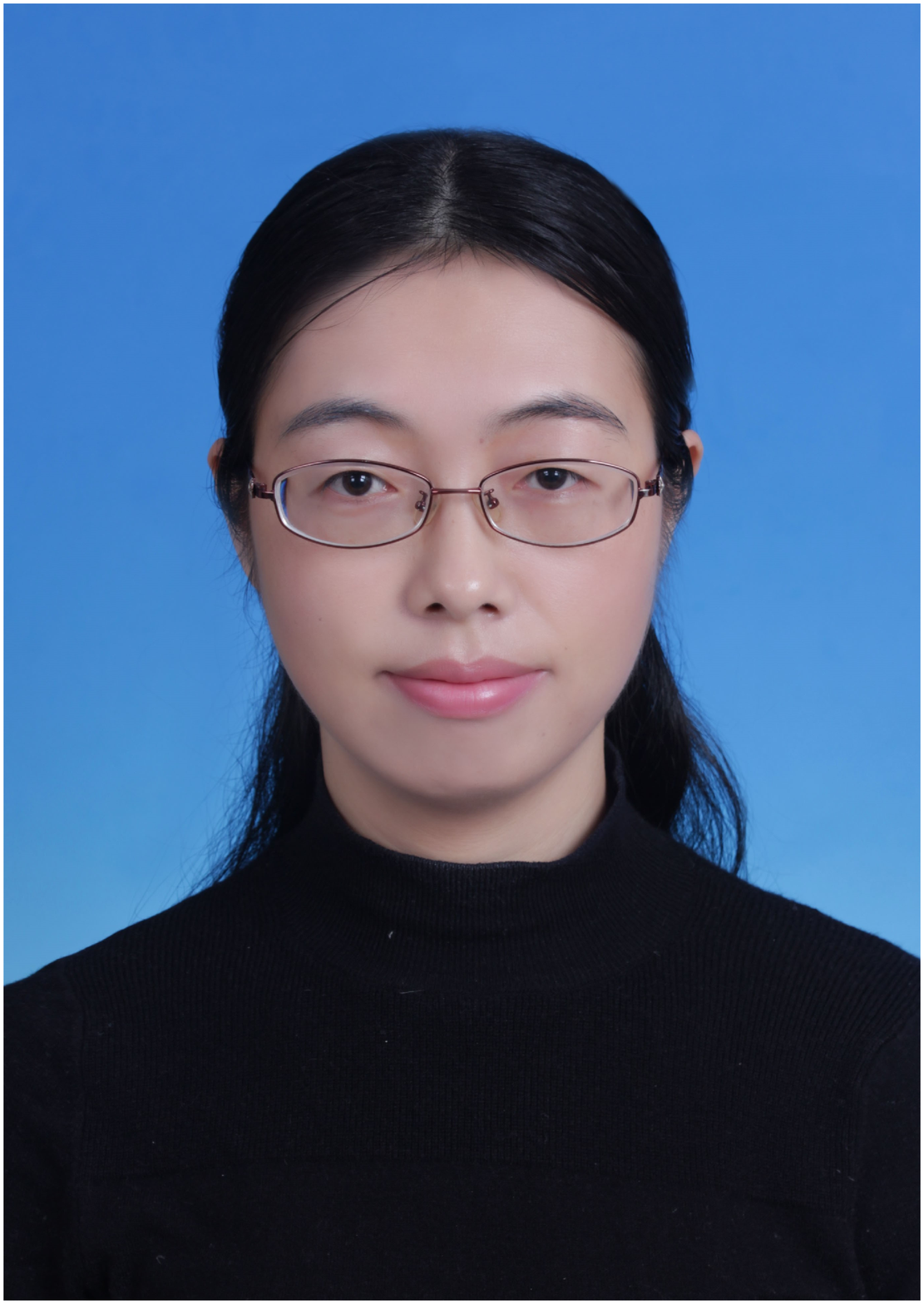}}]{Haiyan Guo} is an Assistant Professor at the Nanjing University of Posts and Telecommunications (NUPT), Nanjing, China. She received her B.Eng. and Ph.D. degrees in signal and information processing from NUPT, Nanjing in 2005 and 2011, respectively. From 2013 to 2014, she was a post-doctoral research fellow with Southeast University. Her research interests include physical-layer security and speech signal processing. \end{IEEEbiography}

\end{document}